\documentclass{article}
\pdfoutput=1

\usepackage[final]{neurips_2019}
\usepackage[ruled]{algorithm2e}
\usepackage[utf8]{inputenc} 
\usepackage[T1]{fontenc}    
\usepackage{hyperref}       
\usepackage{url}            
\usepackage{booktabs}       
\usepackage{amsfonts}       
\usepackage{nicefrac}       
\usepackage{microtype}      
\usepackage{color}
\title{Proximal Policy Optimization for Improved Convergence in IRGAN}

\author{%
  Moksh Jain  \\
  Department of Information Technology\\
  National Institute of Technology Karnataka\\
  Surathkal, India 575025 \\
  \texttt{16it221.moksh@nitk.edu.in} \\
   \And
   Sowmya Kamath S \\
  Department of Information Technology\\
  National Institute of Technology Karnataka\\
  Surathkal, India 575025 \\
   \texttt{sowmyakamath@nitk.edu.in} \\
}

\begin{document}

\maketitle

\begin{abstract}
IRGAN is an information retrieval (IR) modeling approach that uses a theoretical minimax game between a generative and a discriminative model to iteratively optimize both of them, hence unifying the generative and discriminative approaches. Despite significant performance improvements in several information retrieval tasks, IRGAN training is an unstable process, and the solution varies largely with the random parameter initialization. In this work, we present an improved training objective based on proximal policy optimization objective and Gumbel-Softmax based sampling for the generator. We also propose a modified training algorithm which takes a single gradient update on both the generator as well as discriminator for each iteration step. We present empirical evidence of the improved convergence of the proposed model over the original IRGAN and a comparison on three different IR tasks on benchmark datasets is also discussed, emphasizing the proposed model's superior performance.
\end{abstract}

\section{Introduction}
Approaches in information retrieval fall into two broad categories - \textit{Generative models}, which attempt to model the underlying stochastic generative process between the documents and information need (in form of a query) $q \rightarrow d$, and, \textit{Discriminative models}, that predict a relevance score or ranking for a given document-query pair $q+d \rightarrow r$. IRGAN \citep{wang2017irgan} is model which combines the generative and discriminative approaches in information retrieval with a minimax game, in the broad Generative Adversarial Network \citep{goodfellow2014generative} framework. The model achieved state of the art performance on several information retrieval tasks.


Our work improves upon IRGAN by reformulating the optimization objective of the generator based on proximal policy optimization \citep{schulman2017proximal} and incorporating the Gumbel-Softmax reparamterization trick\citep{jang2016categorical} for sampling from the generator. This is coupled with a modified training algorithm for iterative optimization of both the models. We also provide empirical evidence that our proposed approach converges to closer approximates of the Nash Equilibrium for the minimax game, demonstrated by improved performance in several key information retrieval tasks.

\section{Improved Convergence in IRGAN}
The general information retrieval problem can be described as follows: Given a query set $\{q_i\}$, $i \in 1\dots N$, where a set of relevant labelled documents exist for a given query $q_i$ (which is a representation of the actual information need), and a set of documents $\{d_j\}$, $j \in 1\dots M$, the goal is to find the subset of relevant documents for each query in the query set \citep{Manning:2008:IIR:1394399}. Queries can be in the form of search keywords, questions, or even user profiles, while, the documents can be text documents, web links, or answers. The true relevance distribution can be denoted as a conditional probability distribution $p_{true}(d|q, r)$ which represents the relevance of a document $d$ for a given query $q$ and ranking scheme $r$.

\cite{wang2017irgan} proposed the IRGAN model which consists of a generative model $p_\theta(d| q,r)$, which learns to approximate the true relevance distribution $p_{true}(d|q, r)$ and a discriminative model $f_\phi(q,d)$ which learns to approximate the ranking score for a given query-document pair. During training, we sample documents from the generator which the discriminator learns to distinguish from the samples from the true relevance distribution. Here the generative model acts as an adversary for the discriminative model and vice versa, resulting in a zero-sum minimax game between the two models. We describe our proposed model in the following sections. To focus on the proposed changes we limit the discussion to pointwise models, however the arguments can be extended to the pairwise case.

\paragraph{Discriminative Model}
The objective of the discriminator is to maximize the log-likelihood of correctly classifying the documents sampled from the true relevance distribution and the documents sampled from the generator's learned relevant distribution. The discriminator learns a ranking function $f_\phi(q, d)$ to score a given query document pair. This score indicates the relevance of a document for the given query. The output of the discriminator essentially indicates the probability of the given document being sampled from the true relevance distribution given the query. The output of the discriminator can be computed as the sigmoid of the scoring function as follows:
\begin{equation}
D(d|q) = \frac{\exp (f_\phi(q, d))}{1 + \exp(f_\phi(q, d))}
\end{equation}

The output of the discriminator $D$ is expected to be $1$ for samples from the true relevance distribution and $0$ for samples from the estimated distribution. At equilibrium, however, the generator ideally has estimated the exact true relevance distribution, and thus the discriminator should output $\frac{1}{2}$ for all inputs $(q, d)$. The optimal parameters for the discriminator scoring function can be represented as:

\begin{equation}
\phi^* =  \arg \max_\phi \sum_i^N(\mathbf{E}_{d \sim p_{true}(d
| q, r)}[\log(\sigma(f_\phi(d, q_i)))] + \mathbf{E}_{d \sim p_\theta(d|q_i, r)}[\log(1 - \sigma(f_\phi(d, q_i)))])
\end{equation}

\paragraph{Generative Model}
The goal of the generator is to estimate the true relevance distribution, using which it can generate samples which are misclassified by the discriminator. That is, the generator acts as an adversary to the discriminator. Similar to the discriminator, the generator also learns a scoring function, $g_\theta(q, d)$ which reflects the chance of $d$ being selected for the given $q$. We define the generative model from the scoring function using the gumbel-softmax reparametrization trick \citep{jang2016categorical}, as shown in Eq. (3), where, $v_1 \dots v_M$ are i.i.d samples from $Gumbel(0, 1)$, and $\tau$ is the softmax temperature.

\begin{equation}
p_\theta(d_i|q, r) = \frac{\exp((\log g_\theta (q, d_i) + v_i) / \tau)}{\sum_{k=1}^M \exp((\log g_\theta (q, d_k) + v_k) / \tau)}
\end{equation}

The problem of training this generative model can be formulated as a single-step reinforcement learning problem. The generator can be modeled as an agent, whose discrete action space is defined by the documents in the collection, and state space is defined by the queries. The goal is to maximize the reward, which is governed by the output of the discriminator. \cite{wang2017irgan} use a basic policy gradient formulation and use the REINFORCE algorithm \citep{Williams:1992:SSG:139611.139614} to train this agent.

We propose using Proximal Policy Optimization \citep{schulman2017proximal} for training this agent. PPO provides significant improvements over REINFORCE in a number of reinforcement learning tasks. We use PPO with clipped objective function which provides better performance with a relatively straightforward implementation. For incorporating PPO, we define an additional scoring function $g_{\theta'}(q, d)$ whose parameters $\theta'$ are set to  $\theta$ after every $k$ iterations. This additional scoring function can then define a target distribution $p_{\theta'}(d|q, r)$, which functions as a target actor in PPO. The objective function $J^G$ for the generative model can thus be defined as follows:
\begin{equation}
J^G(q_i) = \mathbf{E}_{d \sim p_{\theta'}(d|q_i, r)}[ \min(r_i(\theta)A_i^{p_{\theta'}(d|q,r)}, clip(r_i(\theta), 1 + \epsilon, 1 - \epsilon)A_i^{p_{\theta'}(d|q,r)})]
\end{equation}
\begin{equation}
r_i(\theta) = \frac{p_\theta(d|q,r)}{p_{\theta'}(d|q,r)}
\end{equation}
\begin{equation}
A_i^{p_{\theta}(d|q, d)} = \log (1 + \exp (f_\phi(q, d)))  - \mathbf{E}_{d \sim p_{d|q, r}}[\log(1 + \exp (f_\phi(q, d)))]
\end{equation}
\begin{equation}
\theta^* =  \arg \max_\theta \sum_i^N J^G(q_i)
\end{equation}


The training procedure proposed by \cite{wang2017irgan} involves some pre-training of the generative and discriminative models on the training data. This is followed by an iterative process consisting of several epochs of training of the generator followed by several epochs of training of the discriminator. We discover that this training procedure is highly unstable and prone to getting stuck at the local optimum. We suspect that this might be due to each model being trained against a stationary adversary for several epochs. The procedure is also highly sensitive to the choice of the random seed for initialization of the parameters.

We propose a modified training procedure for training IRGAN (Algorithm 1). We proposed training both models simultaneously within the same epoch. That is, for a single training batch, we update the model parameters for the generator and discriminator within the same iteration. This ensures that model the models have knowledge of the same data at any step, ensuring that they are equally matched. As discussed in the next section, this procedure drastically improves the training stability and also results in faster convergence.

\begin{algorithm}[H]
\SetAlgoLined
\KwResult{Optimal parameters $\phi$ and $\theta$}
 Randomly initialize parameters $\theta$ and $\phi$\;
 Set $\theta' = \theta$\;
 Prepare training data $S$\;
 \Repeat {convergence} {
  Sample $K$ documents for each query $q$ from generator $p_\theta(d|q,r)$\;
  Update $\theta$ with the new PPO formulation\;
  Generate negative samples from $p_\theta(d|q,r)$ for discriminator $f_\phi(q, d)$\;
  Combine generated samples with positive samples from \;
  Update $\phi$ with on the constructed batch\;
  \If{iteration \% k = 0}{
    Set $\theta' = \theta$\;
  }
  }
 \label{algo}
 \caption{Proposed training algorithm for IRGAN}
\end{algorithm}

Given the true relevance distribution, it can also be proved that there exists a Nash equilibrium for this minimax game \citep{goodfellow2014generative}. At equilibrium, the generative model estimates the true relevance distribution ($p_\theta(d| q,r) = p_{true}(d| q,r))$ and the discriminator is unable to distinguish the generated samples from the true samples($D(d|q) = \frac{1}{2}$). In most scenarios however, we do not have the true relevance distribution. We find that our proposed method the generative model learns a closer estimate of the true relevance distribution, leading to improved discriminator performance due to a stronger adversary. This essentially results in the underlying minimax game converging to a closer approximate of the Nash Equilibrium. In the following section, we provide empirical evidence for the same.

\section{Experiments and Results}
We reproduce the results presented by \cite{wang2017irgan} in PyTorch \citep{paszke2017automatic} and then modify the implementation according to the proposed model and run the same experiments on the proposed model with some ablation studies on three information retrieval tasks - Web Search, Item Recommendation and Question-Answering. All the scores presented are the mean values for 5 separate runs with different random seeds.

\paragraph{Web Search}
We use the MQ2008-semi (Million Query track) collection in LETOR 4.0 \citep{qin2010letor} for our experiments in Web Search. Here, the input query is a feature vector containing features extracted from anonymized web search queries and the task is to predict the $id$ of the relevant links. We use the same experimental setup as \cite{wang2017irgan}.

\begin{table}[htpb]
\centering
\caption{Experimental results for Web Search task (MQ2008-Semi Collection)}
\label{tab:wsresult}
\begin{tabular}{l|lll|lll}
 \toprule
   & p@3 & p@5 & p@10 & ndcg@3 & ndcg@5 & ndcg@10 \\ 
 \cmidrule(r){1-7}
 IRGAN & 0.1722 & 0.1653 & 0.1257 & 0.2065 & 0.2225 & 0.2483 \\ 
 IRGAN-PPO & 0.1765 & 0.1678 & 0.1302 & 0.2096 & 0.2267 & 0.2512 \\ 
 IRGAN-SGS & 0.1758 & 0.1692 & 0.1300 & 0.2100 & 0.2256 & 0.2520 \\ 
 IRGAN-SGS+PPO &  \textbf{0.1860} & \textbf{0.1781} & \textbf{0.1384} & \textbf{0.2187} & \textbf{0.2396} & \textbf{0.2619} \\
 \bottomrule
\end{tabular}
\end{table}

\vspace{-0.8em}
\paragraph{Item Recommendation}
For the item recommendation task, we run experiments on the MovieLens-100k dataset \citep{harper2016movielens}. Here, the query is in the form of a user profile and the task is to recommend relevant movies for the given user. The experimental setup remains identical to \cite{wang2017irgan}.

\begin{table}[htpb]
\centering
\caption{Experimental Results for Item Recommendation task (MovieLens-100k collection).}
\label{tab:irresult}
\begin{tabular}{l|lll|lll}
 \toprule
   & p@3 & p@5 & p@10 & ndcg@3 & ndcg@5 & ndcg@10 \\ 
 \cmidrule(r){1-7}
 IRGAN & 0.4072 & 0.3750 & 0.3140 & 0.4222 & 0.4009 & 0.3723 \\ 
 IRGAN-PPO & 0.4131 & 0.3802 & 0.3198 & 0.4301 & 0.4075 & 0.3892 \\ 
 IRGAN-SGS & 0.4159 & 0.3798 & 0.3226 & 0.4294 & 0.4131 & 0.3851 \\ 
 IRGAN-SGS+PPO &  \textbf{0.4227} & \textbf{0.3910} & \textbf{0.3378} & \textbf{0.4486} & \textbf{0.4253} & \textbf{0.4026} \\
 \bottomrule
\end{tabular}
\end{table}

\vspace{-0.8em}
\paragraph{Question Answering}
We consider a specific question answering problem, where, the question is the query and the task is to predict the relevant answers from the given collection. The InsuranceQA dataset \citep{feng2015applying} was used for experiments and the experimental setup remains identical to \cite{wang2017irgan}. We use the pairwise IRGAN formulation in this case.

\begin{table}[htpb]
\centering
\caption{Experimental results for Question-answering task (InsuranceQA collection).}
\label{tab:qaresult}
\begin{tabular}{l|l|l}
 \toprule
   & p@1 (Test 1) & p@1 (Test 2) \\ 
 \midrule
 IRGAN & 0.6444 & 0.6111 \\ 
 IRGAN-PPO & 0.6671 & 0.6397\\ 
 IRGAN-SGS & 0.6733 & 0.6432\\ 
 IRGAN-SGS+PPO &  \textbf{0.7165} & \textbf{0.6784}\\
 \bottomrule
\end{tabular}
\end{table}

\autoref{tab:wsresult}, \autoref{tab:irresult}, \autoref{tab:qaresult} summarise the results of our experiment on the proposed model. IRGAN-PPO consists of only the updated generator objective, IRGAN-SGS incorporates only the improved training procedure, while IRGAN-SGS+PPO utilizes both the proposed modifications. These models are used to evaluate each proposed modification independently. We observe that the precision score of the models with individual and combined modification is $6-11\%$ higher than IRGAN, which indicates that the model learns a better approximation of the true relevance distribution. This claim is also supported by the increased Normalized Discounted Cumulative Gain score, which also indicates the improved performance in graded relevance. The improved graded relevance also indicates the improved of the scoring functions. 

\section{Conclusion} 
In this paper, we presented several improvements to IRGAN in the form of an improved optimization objective for the generator, improved sampling for the generative model and modified training procedure involving single step updating both models. We also present empirical evidence of improved convergence and performance on three different information retrieval tasks. Future work can focus on a more comprehensive theoretical analysis of the effects of the new proposed modifications to the equilibrium of the minimax game.

\bibliography{references}

\end{document}